\documentclass[aps,pra,longbibliography,twocolumn,groupedaddress]{revtex4-2}

\usepackage{amsmath}
\usepackage{amsthm}
\usepackage{amssymb}
\usepackage{multirow}
\usepackage{graphicx,hyperref,bbm,enumitem,bm}  
\usepackage[pdftex]{color}

\usepackage[english]{babel}

\usepackage{xcolor}
    \usepackage{float}

    \usepackage[toc,page]{appendix}
    \def\ii{{\rm i}}
    \def\tr#1{{\rm tr}{#1}}
    \def\1{\mathbbm{1}}

    \def\lph{\lambda_{\mathrm{ph}}}
    \def\l2{\lambda_2}
    \def\leff{\lambda_{\mathrm{eff}}}
    \def\lp{\lambda_{\mathrm{ps}}}

\begin{document}
        
        \title{Arbitrary relaxation rate under non-Hermitian matrix iterations}
        
        \author{Ja\v s Bensa}
        \affiliation{Department of Physics, Faculty of Mathematics and Physics, University of Ljubljana, 1000 Ljubljana, Slovenia}
        
        \date{\today}
        
        \begin{abstract}
            We study the exponential relaxation of observables, propagated with a non-Hermitian transfer matrix, an example being out-of-time-ordered correlations (OTOC) in brickwall (BW) random quantum circuits. Until a time that scales as the system size, the exponential decay of observables is not usually determined by the second largest eigenvalue of the transfer matrix, as one can naively expect, but it is in general slower -- this slower decay rate was dubbed ``phantom eigenvalue''. Generally, this slower decay is given by the largest value in the pseudospecturm of the transfer matrix, however we show that the decay rate can be an arbitrary value between the second largest eigenvalue and the largest value in the pseudospectrum. This arbitrary decay can be observed for example in the propagation of OTOC in periodic boundary conditions BW circuits. To explore this phenomenon, we study a 1D biased random walk coupled to two reservoirs at the edges, and prove that this simple system also exhibits phantom eigenvalues.
        \end{abstract}
        
        \maketitle

\textbf{\textit{Introduction}} -- Physical phenomena are often considered intriguing because they stand counter to our classical intuition. Quantum mechanics offers a plethora of such phenomena, for example the uncertainty principle \cite{Heisenberg1927} or quantum entanglement \cite{EPR}, phenomena which can be considered ``spooky''  when compared to classical expectations. Even though quantum mechanics is \textit{the} counter-intuitive theory, nonetheless familiarity has been developed for closed systems, which are subject to Hermitian evolution. However, real life systems are never closed, leading to non-Hermitian propagation, for example the Linblad master equation \cite{breuer2002}. Non-Hermitian physics is often thought of as counter-intuitive even among quantum mechanics itself, and leads to interesting behaviour such as non-Hermitian skin effect \cite{Lee16,Torres18,Wang,Shen18,Kunst18,Lee19,Kawabata19,Gong20,Slager20,Okuma20,Zhang20,Kunst21,Ueda20,Torres20,Okuma22,Lee23}, relaxation in Liouvillian systems \cite{linblad1,linblad2,linblad3,linblad4,linblad5} and phantom eigenvalues \cite{PRX,PRR,marko22,PRA,marko23,zhou23}. In this paper, we delve into the phantom eigenvalue phenomenon, which appears in the decay of observables propagated using a non-Hermitian transfer matrix.

Observables $O(t)$ which are propagated using a transfer matrix tend to decay towards their asymptotic value $O(\infty)$ exponentially
\begin{equation}
    O(t) - O(\infty) \propto \leff^t(t).
\end{equation}
In recent work \cite{markov,oliveira,marko08,PRX17,PRR} it was shown that purity and out-of-time-ordered correlations (OTOC) \cite{lashkari13,shenker14,maldacena16} in random circuits can be propagated using a transfer matrix approach. It was noted \cite{PRX,PRR} that these quantities exhibit a two stage exponential relaxation towards their asymptotic values, namely
\begin{equation} 
    \leff(t) = \begin{cases}
                    \lph, \quad t \lesssim n \\
                    \lambda_2, \quad t \gtrsim n 
                \end{cases}
\end{equation}
where $\lambda_2$ is the second largest eigenvalues of the transfer matrix (the largest is equal to $1$ and corresponds to the stationary state). Note that the initial decay $\lph$ persists until extensive times in the system size, meaning that the effective decay $\lph \neq \lambda_2$ will be present for all times in the thermodynamic limit. This behavior was dubbed phantom eigenvalue in previous papers \cite{PRX}.

Typically, the decay in processes involving a non-Hermitian transfer matrix is determined by the largest value $\lp$ in the pseudospectrum \cite{marko22,PRA}. In this paper we instead observe that the decay can be an arbitrary value between $\lambda_2$ and $\lp$. This happens for example in the OTOC relaxation in brickwall (BW) periodic boundary conditions (PBC) random quantum circuit. We work out the details behind this arbitrary relaxation and find that, surprisingly, the behaviour is not determined exclusively by the properties of the transfer matrix, but depends on the specifics of initial vectors used in the iteration.

Such a decay was observed by examining the behavior of OTOC in a chain of $2n$ qudits of dimension $q$

\begin{equation}
    O(t) = 1- \frac{1}{2^{2n}} \tr ( X_i(t) Y_j X_i(t) Y_j ) ,
    \label{eq:app-OTOC}
\end{equation}
where $X_i$ and $Y_j$ are two local Pauli-like matrices located at qudit $i$ and $j$, respectively. Note that $O(t)$ depends on both $i$ and $j$, but we will leave the dependence implicit. The matrix $X_i$ is propagated in time with the BW random quantum circuit $U$ of depth $t$ ($2t$ rows of a BW geometry), namely $X_i(t) = U^\dagger X_i U$. The random matrix $U$ is composed of two sites independent Haar unitaries. Because the unitaries in the circuit are random, $O(t)$ is independent on the choice of the matrices $X$ and $Y$. 

\begin{figure}[h]
        \begin{center}
        \includegraphics[width=70mm]{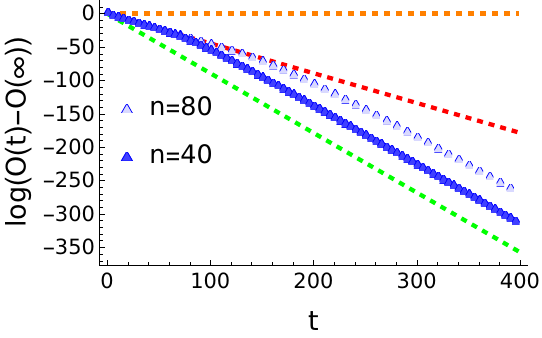} \caption{OTOC in BW PBC circuit initially decays as $\lph=(4/5)^2$ (red slope), which is between $\lambda_2=(4/5)^4$ (green slope) and $\lp=1$ (red slope). With increasing system size the transition point from $\lph$ to $\lambda_2$ grows linearly, making the decay $\lph$ the true decay of OTOC in the TDL.}
    \label{fig:PBC_intro}
    \end{center}
\end{figure}

First, we will look into PBC random quantum circuits. Specifically, we will derive a transfer matrix propagation approach to study the OTOC. Consequently, we observe an intriguing behavior in its decay towards its asymptotic value. Remarkably, the decay in the thermodynamic limit (TDL) is neither given by the second largest eigenvalue of a transfer matrix, nor by the pseudospectrum, but
\begin{equation}
    \lambda_2 < \lph < \lp.
\label{eq:lph_vmes}
\end{equation}
See Fig.~\ref{fig:PBC_intro} for a graphic representation.

To understand better this phenomenon, we will delve in the simpler case of OTOC in BW open boundary condition (OBC) circuits . Surprisingly, we will find that the transfer matrix used to describe OTOC dynamics can be also used to describe other processes, such as a biased 1D random walk with dissipation at the edges, see App.~\ref{app:OBC-randomwalk}. In App.~\ref{app:OBC-randomwalk}, we leverage the equivalence between OTOC dynamics and a biased random walk to obtain a closed form solution for $O(t)$ when $j=1$. Focusing on general iterations of this transfer matrix, we will explain why $\lph$ can be a value between $\lambda_2$ and $\lp$.

\textbf{\textit{Periodic boundary conditions}} -- The average time evolution of OTOC in BW PBC circuits is equivalent to a partition function of an Ising-like model \cite{adam18}. Shortly, the partition function is obtained by summing certain domains of up-spins that evolve on a 1D Ising chain of $n$ sites \cite{chain}, one site for each random gate, see App.~\ref{app:adamPRX} for a brief overview. In App.~\ref{app:PBC-der} we derive a Markovian evolution of OTOC by composing a transfer matrix that propagates all domains in the previously mentioned 1D Ising chain \cite{zakaj_ne_markov}. To calculate the average OTOC at time $t$, we iterate the transfer matrix $A$ on an initial vector $ |v\rangle $ $t$ times, $t \in \mathbb{N}$. The domains relevant for the partition function from \cite{adam18} are extracted by projecting $A^t |v\rangle $ on the vector $ \langle p|$
    \begin{equation}
        O(t) = \langle p|A^t|v\rangle.
    \label{eq:OTOC_PBC}
    \end{equation}
Note that $ \langle p|$ depends on the qudits $j$.

\begin{figure}[h]
        \begin{center}
        \includegraphics[width=80mm]{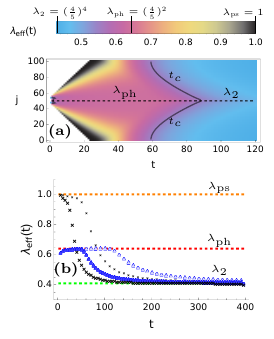}
        \caption{OTOC time dependence for a BW PBC circuit, $q=2$. (a) shows the propagation of $O(t)-O(\infty)$ for different final positions $j$ for $i=n$, $n=50$. The gray curve denotes the critical time $t_c$ when $\leff = 0.55$. (b) shows the iteration of $A$ on two different sets of initial vectors: blue symbols denote the choice of vectors that give OTOC dynamics (black dashed line from (a)), black symbols denote the choice of the physical $ \langle p|$, but $p_{n(n-1)+1}=0$, and random $v_k$, $\sum_k v_k = 1$, that decays as $\lp^t$. The green dotted line denotes the decay given by $\lambda_2$, the red dotted line denotes $\lph$ and the orange one the decay for $\lp$. (b) also compares data for $n=80$ (light symbols) with data for $n=40$ (dark symbols), which shows that the initial decays persist until $t \sim n$.}
    \label{fig:PBC}
    \end{center}
\end{figure}

Once we obtain a transfer matrix approach for computing OTOC, we can compare the actual exponential decay with the second largest eigenvalue $\lambda_2$ of $A$. Fig.~\ref{fig:PBC} illustrates that the actual decay $\lph$ of OTOC between qubits $n$ and $j=1$, $q=2$, lies between $\lambda_2=(4/5)^4$ (computed in App.~\ref{app:PBC-diag}) and $\lp=1$, namely $\lph = (4/5)^2$ \cite{lambda_2_OTOC-purity}. Exact results were computed also for general $q$, namely $\lambda_2 = 16 q^4/(1+q^2)^4$ (see App.~\ref{app:PBC-diag}), $\lp=1$ and $\lph = \lph^2$. This intriguing behavior is retained for all possible choices of $q$ and $j$. The underlying reason for this behavior can be attributed to our specific selections of $ \langle p|$ and $ |v \rangle$. For general or randomly chosen initial vectors, the expression $ \langle p|A^t|v\rangle$ decays as $\lp^t$ until $t \sim n$, see Fig.~\ref{fig:PBC}b for an example.

\textbf{\textit{Open boundary conditions}} -- Let us continue by examining the propagation of OTOC for BW OBC random quantum circuits. Similar to the case of PBC, OTOC can be obtained by evolving all domains on a 1D Ising chain of $n$ sites and summing only the relevant domains that are described in \cite{adam18,PRR} or in App.~\ref{app:adamPRX}. To analyze the correlation between the first and the $j$-th qudit, we focus just on the right edge of our domain since the left edge remains fixed at the left boundary of the system. Consequently, at every time there are only $n$ possible domains, one for each domain width, in contrast to the $\approx n^2$ domains in the PBC scenario. The initial vector $ |v\rangle$, $v_k = \frac{q^4}{q^4-1} \delta_{1,k}$, is propagated with the transfer matrix

\begin{equation}
    A = \begin{pmatrix}
            \multicolumn{4}{c}{\multirow{3}{*}{T}} & 0 \\
            \multicolumn{4}{c}{} & \vdots \\
            \multicolumn{4}{c}{} & 0 \\
            0 & \dots & 0 & \sigma & 1
        \end{pmatrix};
    \quad
    T = 
    \begin{pmatrix}
        \delta & \tau & 0 & \dots & 0  \\
        \sigma & \delta & \tau & \dots & 0  \\
        \vdots & \ddots & \ddots & \ddots &  \vdots  \\
        0 & \dots & \sigma & \delta & \tau   \\
        0 & \dots & 0 & \sigma & \delta   \\
   \end{pmatrix},
\label{eq:transfer_matrix}
\end{equation}
where $\delta = \frac{2 q^2}{(1+q^2)^{2}}$, $\tau=\frac{1}{(1+q^2)^{2}}$ and $\sigma=\frac{q^4}{(1+q^2)^{2}}$. See App.~\ref{app:OBC-der} for more details. Note that the transfer matrix $A$ exhibits a distinct structure. Namely, the first $n-1$ components are propagated using a tridiagonal Toeplitz matrix $T$ \cite{toeplitz}. If we are interested in the OTOC between the first qudit and the $j$-th one, the summation over relevant domains is represented with the inner product with the vector $ \langle p|$, $p_k = 1$ if $k \geq j$ and 0 otherwise, leading to $O(t) = \langle p|A^t|v\rangle$.

In the physical systems that we analyze, the vector $ |v\rangle $ is localized at the leftmost position. Thus, for times $t<n$, the iteration with $T$ is equivalent to the iteration with $A$. Therefore, when we want to analytically study the iteration until extensive times $t \sim n$ (the interesting domain) we can focus on the properties of the iteration $O(t) = \langle p|T^t|v\rangle $. However, note that $O(\infty)$ is still defined as $\lim_{t \rightarrow \infty} \langle p|A^t|v\rangle $.

Before moving to the properties for general $\delta$,$\tau$ and $\sigma$, we will first examine the classical limit $q \rightarrow \infty$. In this case, only $\sigma \neq 0$ and $T$ corresponds to a transposed Jordan kernel. We will generalize this transposed Jordan kernel to a transposed Jordan block with diagonal elements equal to $\delta$ (which is zero in for $q \rightarrow \infty$) and lower diagonal elements $\sigma$. Note that the $n$ times degenerate eigenvalue of $T$ is $\delta$ and the largest value in the pseudospectrum is $\delta+\sigma$. The physical case $q \rightarrow \infty$ is recovered by setting $\delta=0$ and $\sigma=1$. When $ |v\rangle $ is left-localized, the iteration for times $t < n$ can be expressed as follows

\begin{align}
    O(t) &= \langle p | T^t | v \rangle  \nonumber \\
         &= \sum_{r=0}^{\min(t,n-1)} \binom{t}{r} \delta^{t-r} \sigma^r \sum_{j=1}^{n-r} \langle p |e_{j+r} \rangle \langle e_{j} | v \rangle \nonumber \\
         &= \sum_{r=0}^{\min(t,n-1)} \binom{t}{r} \delta^{t-r} \sigma^r C(r),
\label{eq:jordan_spectral}
\end{align}
The second row is obtained by replacing the $t$-th power of the Jordan block with the $t$-th power of a diagonal matrix of entries $\delta$ plus the nilpotent matrix with lower diagonal elements equal to $\sigma$. The vectors $e_j$ with $k$-th component $\delta_{j,k}$ are generalized eigenvectors of the nilpotent matrix. The quantity $C(r)$ from the last row is the convolution between the initial vectors $ \langle p|$ and $ |v\rangle$ . If $C(r)$ is independent of $r$, for example for $p_k = 1$ and $v_k=\delta_{k,1}$, then the sum over $r$ can be evaluated, and $O(t) \propto (\delta+\sigma)^t$ for $t<n$. It is worth noting that $\delta+\sigma$ corresponds to the largest value in the pseudospectrum of $T$. Interestingly, if $C(r)=\mu^{-r}$, for instance when $p_k = \mu^{-k}$ and $v_k = \delta_{k,1}$, then $O(t) = (\delta+\sigma/\mu)^t$ for $t<n$. In the TDL, if we wish to normalize $ \langle p| $, we must choose $ \mu > 1$. In all the examples $O(\infty)=0$, consequently, the decay of $O(t)$ to its asymptotic value becomes an \textit{arbitrary} number between $\delta$, the largest eigenvalue, and $\delta+\sigma$, the pseudospectrum of $T$. This brief derivation illustrates that we cannot determine the decay of $O(t)$ solely by examining the properties of the transfer matrix. In the case $C(r) = \mu^{-r}$, the decay depends on the specific form of the initial vectors.

\begin{figure}[h]
    \begin{center}
    \includegraphics[width=80mm]{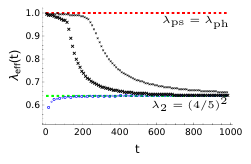}
        \caption{Iteration of $A$ (OBC, $q=2$) on two different sets of initial vectors: blue symbols denote the choice of vectors that give OTOC dynamics, black symbols denote the choice $p_k=1-\delta_{n,k}$ and random $v_k$, $\sum_k v_k=1$, that decays as $\lp^t$. The green dotted line denotes $\lambda_2$, the red dotted line the value $\lp$. The plot also compares data for $n=80$ (small symbols) with data for $n=40$ (big symbols), which shows that the initial decays persist until $t \sim n$.}
    \label{fig:OBC}
    \end{center}
\end{figure}

Now, let us move back to the case of general $q$. The largest value $\lp=1$ in the pseudospectrum of $T$ is larger than its largest eigenvalue $\lambda_2 = \delta+2 \sqrt{\sigma \tau}$, see App.~\ref{app:OBC-properties} for the properties of $T$. We would expect that quantities like $ O(t)-O(\infty) = \langle p|T^t|v\rangle$ decay as $\lp^t$. However, it is known that OTOC decays as $\lambda_2^t$ \cite{PRR}. It turns out that the initial vectors for OTOC are special, see App.~\ref{app:OBC-randomwalk} for an argument made by considering OTOC dynamics as a biased random walk or App.~\ref{app:OBC-sums} for an exact computation of $O(t)$. When considering general (or random) $ \langle p| $ and $ |v \rangle $, the decay of $O(t)$ would indeed be given by the largest value in the pseudospectrum, see Fig.~\ref{fig:OBC}.

\begin{figure}[h]
    \begin{center}
    \includegraphics[width=70mm]{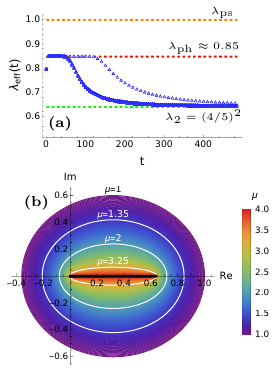}
        \caption{(a) Iteration of $A$ (OBC, $q=2$) for $p_k=\mu^{-k}$ and $v_k=\delta_{k,1}$, $\mu=1.35$. The quantity $ \langle p|A^t|v\rangle$ decays towards its asymptotic values as $\lph^t \approx 0.85^t$, and $\lambda_2 < \lph < \lp$. The red line denotes $\lph$, the orange line the value $\lp > \lph$ and the green line the asymptotic decay of finite system sizes $\lambda_2$. The plot shows data for two system sizes, namely $n=80$ for light symbols and $n=40$ for dark symbols, which shows that $\lph$ persists until times extensive in the system size. (b) Cartoon representation of the pseudospectrum of the matrix from Eq.~\ref{eq:tridiagonal_rescaled}. For different exponential localization $\mu$ of the initial vector we get different pseudospectra. As we increase $\mu$, we get from the pseudospectrum of $A$ for $\mu=1$ to the spectrum of $A$ for $\mu=4$. The solid black line on the real axis corresponds to the spectrum of the matrix.}
    \label{fig:arbitrary}
    \end{center}
\end{figure}

\textbf{\textit{Arbitrary decay}} -- We have observed that the decay $\lambda_2^t$ is achieved through the use of special vectors. Is it possible to get an arbitrary decay rate with an appropriate choice of $ \langle p|$ and $ |v\rangle $, as we saw for Jordan blocks? Fig.~\ref{fig:arbitrary}a) shows how $O(t)-O(\infty) = \langle p|A^t|v\rangle$ decays for $v_k=\delta_{k,1}$ and $p_k = \mu^{-k}$, $\mu = 1.35$. Although the largest value in the pseudospectrum is $1$, this quantity decays as $\approx 0.85^t$, which is between $\lambda_2$ and $\lp$.

The arbitrary decay can be intuitively understood in terms of the pseudospectrum. Instead of propagating $O(t) = \langle p|T^t|v\rangle $ for $p_k = \mu^{-k}$ and $v_k = \delta_{k,1}$, we instead express it like $ \langle \tilde{p}|D^{-1} T^t D|\tilde{v}\rangle$, where $\tilde{p}_k = 1$, so $D$ is a diagonal matrix with diagonal elements $D_{k,k}=\mu^k$. The action of $D^{-1}$ retains the right vector's form but rescales it. Recall that non-unitary similarity transformations can alter the matrix's pseudospectrum \cite{trefethen}. In fact, the largest value in the pseudospectrum of

\begin{equation}
    D^{-1} T D = \begin{pmatrix}

        \delta & \tau \mu & 0 & \dots & 0 \\
        \frac{\sigma}{\mu}& \delta & \tau \mu & \dots & 0 \\
        \vdots & \ddots & \ddots & \ddots & \vdots \\
        0 & \dots & \frac{\sigma}{\mu} & \delta & \tau \mu \\
        0 & \dots & 0 & \frac{\sigma}{\mu} & \delta
        \end{pmatrix}
\label{eq:tridiagonal_rescaled}
\end{equation}
is $\lp = \delta + \frac{\sigma}{ \mu} + \tau\mu $ which coincides with the decay of $O(t)$ from Fig.~\ref{fig:arbitrary}a for $\mu=1.35$. A schematic representation of the pseudospectrum of $D^{-1} T D$ for different values $\mu$ is shown in Fig.~\ref{fig:arbitrary}b. It turns out that $\lph$ can only be between $\lambda_2$ and $\lp$. To see this one must solve exactly all the sums in the expression $O(t) = \langle p|T^t|v\rangle$, see App.~\ref{app:OBC-sums} for the computation of $\lph$. We conclude that the decay of $O(t)$ towards its asymptotic value is not determined solely by the properties of the transfer matrix, but it is highly dependent on the initial vectors used in the iteration. 

Note that with Eq.~\ref{eq:tridiagonal_rescaled}, we could get a non-Hermitian matrix from an initially Hermitian matrix ($\sigma=\tau$). Does this mean that a decay slower than $\lambda_2^t$ is possible in Hermitian systems? It turns out that achieving $\lph > \lambda_2$ is only possible if we select $p_k = \mu^{-k}$ with $\mu < 1$. In the TDL, this implies that the vector $ \langle p|$ is not normalizable, causing $O(t)$ to be exactly equal to zero already at $t=0$, see App.~\ref{app:herm} for the proof. Since the OTOC is exactly zero in the TDL, $\lph > \lambda_2$ in Hermitian systems becomes a finite size effect. 

\textbf{\textit{Discussion}} -- We studied the decay of observables $O(t) = \langle p|A^t|v\rangle $ propagated with a non-Hermitian transfer matrix $A$. Such systems can be found when studying purity or OTOC propagation in random quantum circuits or other Markovian systems. The peculiarity of these systems is that they can exhibit phantom eigenvalues, that is, the convergence of $O(t)$ towards its asymptotic value is not determined by the second largest eigenvalue $\lambda_2$ of $A$ but rather by the largest value $\lph$ in the pseudospectrum of $A$. In this paper we have shown that this is not always the case. The exponential rate at which $O(t)$ relaxes can be an arbitrary value between $\lambda_2$ and $\lph$. Such ``arbitrary'' decay happens in physical systems, for example in OTOC relaxation in PBC BW random quantum circuits. To compute the actual decay rate of $O(t)$ one must not just look at properties of the transfer matrix $A$, but rather at the whole system. Namely, we found that when $ \langle p|$ was exponentially localized, this changed the convergence rate of $O(t)$ to an arbitrary value, which depends on the localization length of $ \langle p|$.

Although the decay is not solely determined by the transfer matrix properties, the pseudospectrum still plays a crucial role in the computation of the phantom decay $\lph$. However, it is not the pseudospectrum of the transfer matrix $A$ which we have to explore, but rather the pseudospectrum of the transformed matrix $ D^{-1} A D$, where $D$ is chosen so that $ \langle p| D$ is exponentially localized. Currently, it is understood that the pseudospectrum of $A$ (or $D^{-1} A D$) determines the exponential relaxation of observables for general initial vectors. However, for special choices of initial vectors the observable can decay as $\lambda_2$. Ultimately, it would be useful to develop a general technique to determine whether an observable decays as $\lph > \lambda_2$ solely from looking at properties of $A$ and initial vectors. This will be ground for future studies.

\textbf{\textit{Acknowledgements}} --  I would like to thank M.~\v Znidari\v c, J.~M.~Bhat and K.~Kavanagh for valuable discussions and comments on the manuscript. Support from Grants No. P1-0402 and J1-4385 from the Slovenian Research Agency is acknowledged.

\bibliography{U4_OTOC}

\begin{thebibliography}{47}
\bibitem [{cha()}]{chain}%
  \BibitemOpen
  \href@noop {} {}\bibinfo {note} {Beware that we have two different chains,
  namely the 1D chain of qudits and the 1D Ising chain. The former has $2n$
  qudits, the latter $n$ sites.}\BibitemShut {Stop}%
\bibitem [{zak()}]{zakaj_ne_markov}%
  \BibitemOpen
  \href@noop {} {}\bibinfo {note} {We can obtain a Markovian propagation of
  OTOC also following the procedures explained in \cite{PRR}. However, the
  resulting transfer matrices in the latter approach are exponentially big in
  the system size, which makes it difficult to access large system sizes. The
  advantage of the Markov chain from \cite{PRR} resides in the fact that it is
  possible to map circuits with non-random two-site gates and random one-site
  gates to a Markov chain; the Markov chain obtained from domain propagation
  can be obtained only for Haar random two-site gates.}\BibitemShut {Stop}%
\bibitem [{lam()}]{lambda_2_OTOC-purity}%
  \BibitemOpen
  \href@noop {} {}\bibinfo {note} {The given value for the phantom eigenvalue
  for BW PBC is exact, because it coincides with the second largest eigenvalue
  for BW OBC as well as with the decay of OTOC in infinite systems. The OTOC
  propagation in all three systems must be identical for times $t\sim n$,
  before the OTOC hits or ``feels'' the boundaries of the system.}\BibitemShut
  {Stop}%
\end{thebibliography}

\onecolumngrid
\appendix

\section{Mapping of OTOC dynamics to the partition function of an Ising model}
\label{app:adamPRX}

In this Appendix, we present the derivations from \cite{adam18} that show that the average OTOC at time $t$ can be expressed as a partition function for a classical grid of Ising spins. Details about this reduction can be found in \cite{adam18}. Here, we shall only explain the general idea. Contrary to \cite{adam18}, where the authors deal with infinite systems, we want to obtain a simple method to propagate OTOC in finite systems with either OBC or PBC. We found that it is possible to construct a Markov chain on a space of dimension $\sim n$ by considering the vertical axis of the grid of spins from \cite{adam18} as time, and propagating the resulting 1D domain of Ising spins in time.

\begin{figure}[h]
    \begin{center}
    \includegraphics[width=70mm]{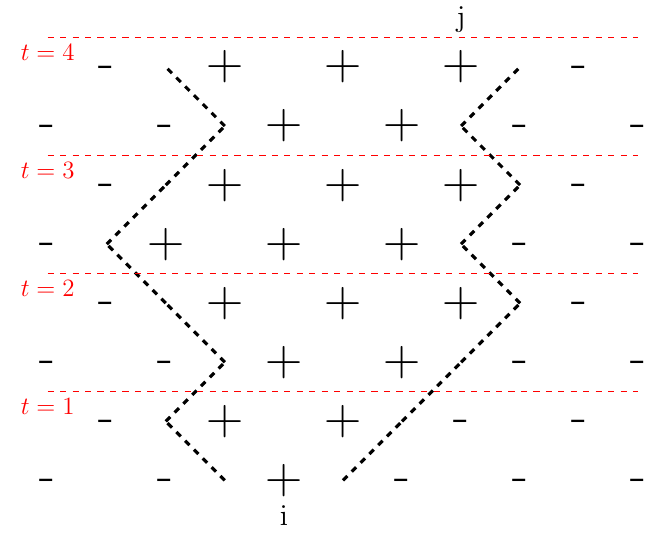}
        \caption{Schematic representation of a domain that must be summed in the partition function for $O_{i,j}(t=4)$. The domain starts around the site of the initial Pauli-like matrix $i$ and contains the site $j$ at time $t=4$.}
    \label{fig:grid}
    \end{center}
\end{figure}

Let us quickly summarize \cite{adam18}. OTOC are 

\begin{equation}
    O(t) = 1- \frac{1}{2^{2n}} \tr ( X_i(t) Y_j X_i(t) Y_j ) ,
    \label{eq:app-OTOC}
\end{equation}
see main text for details. After averaging over all the $2$-site random unitaries in the circuit, $O(t)$ becomes a partition function of a grid of Ising spins with dimensions $n \times 2t$, which is tilted by $45$ degrees, see Fig.~\ref{fig:grid}. The grid is obtained by replacing all the $2$-site unitaries with two-level spins $s \in \{+,-\}$. To obtain $O(t)$, one must sum over all 2D domains that start on the qudit $i$ and contain the qudit $j$ at time $t$, see Fig.~\ref{fig:grid}.

Each domain has its own weight, which is determined by the number of $+$ spins in the top edge of the grid and by length of the domain walls. For each $+$ in the top edge we multiply the weight of the domain by $q^2$. Each opposite neighboring spins (horizontally) contribute with $q/(q^2+1)$. For the example in Fig.~\ref{fig:grid}, the weight of the domain is $(q^2)^3 \cdot (\frac{q}{q^2+1})^{16}$, because the domain has 3 qudits in the top edge and 16 differently oriented neighboring spins. In an infinite system, the number of differently oriented neighboring spins is equal to $4t$, but in a finite system it could be smaller if the domain hits the boundary of the system or if at some point in time the $+$ spin span over all sites.

OTOC in BW circuits can be expressed as a weighted sum of domains on a 2D grid that start at qudit $i$ and contain the qudit $j$ in the top edge. Instead of looking at the partition function of a 2D grid, as the authors of \cite{adam18} did, in the following Sections we will compute OTOC by evolving all possible domains of the 1D Ising chain, obtained by considering the vertical axis of the 2D grid as time. To obtain $O(t)$ we then just sum the weight of relevant domains at time $t$. These weights can be now though as probabilities in this Markovian propagation. These Markovian propagation of domain wall to compute OTOC is preferable when dealing with finite systems, because it is easy to take in account the boundaries of the system. Computing partition functions of the 2D grid of Ising spins works well for infinite systems, but gives complicated solutions expressed with recursion for finite systems \cite{PRR}. Using the Markovian propagation of domain weights described in this paper, in App.~\ref{app:OBC-randomwalk} we will see that we can obtain a simpler close-form solution of OTOC in OBC circuits.

\section{PBC case}
\label{app:PBC}

\subsection{Transfer matrix derivation}
\label{app:PBC-der}

Here we derive a Markovian propagation of all domains for PBC circuit. As described in App.~\ref{app:adamPRX}, OTOC is then recovered by summing only certain domains that contain the site $j$ at time $t$.

To begin, we shall encode all domains in the vector $ |v\rangle $ such that the domain beginning at the $i_z$-th spin and having width $w \in \{1,\dots,n-1\}$ is written in the $n \cdot i_z + w$ component of $ | v\rangle $. Note that $n$ counts the number of spins, which is half of the number of qudits. The only domain of width $n$ will be put in the last position of $ |v\rangle $, so $ | v\rangle $ will be a vector of $n \cdot(n-1)+1$ components.  

To propagate $ |v\rangle $ in time, we can construct a transfer matrix that propagates all domains of the 1D chain of Ising spins (see App.~\ref{app:adamPRX}. Following App.~\ref{app:adamPRX} we get

\begin{equation}
    A = 
        \begin{pmatrix}
            \multicolumn{4}{c}{\multirow{1}{*}{T}} & 0 \\
            \multicolumn{4}{c}{b} & 1 
        \end{pmatrix},
    \label{eq:matrixPBC}
    \end{equation}
where $T$ is a block circulant matrix

\begin{equation}
    T =
    \begin{pmatrix}
        C & U & 0 & \dots & D \\
        D & C & U & \dots & 0 \\
        \vdots & \ddots & \ddots & \ddots & \vdots \\
        0 & \dots & D & C & U \\
        U & \dots & 0 & D & C
    \end{pmatrix}.
    \label{eq:TPBC_circulant}
    \end{equation}
The vector $b$ describes the transition probabilities to the steady state, i.e. a 1D chain of Ising spins with all $+$. $C$ dictates how the domain width changes for fixed $i_z$. $U$ and $D$ describe how to obtain domains with $i_z -1$ or $i_z+1$ from $i_z$, respectively. The matrices $D$, $U$ and $C$ are tridiagonal matrices because of locality in the random walk of domain width and have dimension $(n-1)\times(n-1)$. $T$ has a block circulant form because of the locality in the changes in $i_z$. These matrices are

\begin{align}
    C &=
    \begin{pmatrix}
        3 \tau \sigma & \delta \tau & 0 & \dots & 0 \\
        \delta \sigma & 4 \tau \sigma & \delta \tau & \dots & 0 \\
        \vdots & \ddots & \ddots & \ddots & \vdots \\
        0 & \dots & \delta \sigma & 4 \tau \sigma & \delta \tau \\
        0 & \dots & 0 & \delta \sigma & 3 \tau \sigma
    \end{pmatrix} 
    \\
    D &=
    \begin{pmatrix}
         \tau \sigma & \delta \tau & \tau^2 & 0 & \dots & 0 \\
        0 & \tau \sigma & \delta \tau & \tau^2 & \dots & 0 \\
        \vdots & \ddots &  \ddots & \ddots & \ddots & & \\
        0 & \dots & 0 & \tau \sigma & \delta \tau & \tau^2 \\
        0 & \dots & 0 & 0 & \tau \sigma & \delta \tau \\
        0 & \dots & 0 & 0 & 0 & \tau \sigma \\
    \end{pmatrix} 
    \\
    U &=
    \begin{pmatrix}
        \tau \sigma & 0 & 0 & 0 & \dots & 0 \\
        \delta \sigma & \tau \sigma & 0 & 0 & \dots & 0 \\
        \sigma^2 & \delta \sigma & \tau \sigma & 0 & \dots & 0 \\
        \vdots & \ddots & \ddots & \ddots & \ddots &  \\
        0 & \dots & \sigma^2 & \delta \sigma & \tau \sigma & 0 \\
        0 & \dots & 0 & \sigma^2 & \delta \sigma& \tau \sigma 
    \end{pmatrix}.
    \label{eq:PBCblocks}
\end{align}
where $\delta = \frac{2q^2}{(1+q^2)^2}$, $\tau = \frac{1}{(1+q^2)^2} $ and $\sigma = \frac{q^4}{(1+q^2)^2} $. A row of $T$ is thus composed of $n$ blocks with size $(n-1)\times(n-1)$. As last, the last row $b$ of $A$ is a vector of $n (n-1)$ components. The non-zero components of $b$ are $b_{(n-1)i-1}=\sigma^2$ and $b_{(n-1)i}= \delta \sigma + q^2 \sigma$ for $i \in \{1,\dots,n\}$.

We shall always begin with the initial vector $ |v\rangle $ located on the last spin. This choice is irrelevant because of the periodic boundary conditions. However, we must be careful when we choose $ \langle p|$. Depending on the position of the qudit $j$, the non-zero components are $p_{(n-1)(i-1)+k}=1$ for $i\in \{1,\dots,n\}$ and $k \in \{j-i+1 \mod n-1, \dots,n-1\}$ and $p_{n(n-1)+1}=1$. The vector $ |v\rangle $ is obtained by propagating the domain on the last site for a half-time step. The only nonzero components of $ |v\rangle $ are $v_1 = q^2$, $v_{(n-1)^2+1}=q^2$ and $v_{(n-1)^2}=q^4$.

\subsection{Diagonalization}
\label{app:PBC-diag}

To diagonalize the transfer matrix $A$, one can begin by diagonalizing the block circulant matrix $T$. If $\lambda_k$ and $ |v_k\rangle $ are an eigenpair of $T$, then $\lambda_k$ is also an eigenvalue of $A$ whose eigenvector is obtained from $ |v_k\rangle $ by adding one last component $ \frac{ \langle b|v_k\rangle}{\lambda_k-1} $. There is also an eigenvector of $A$ which is not derived from the eigenvectors of $T$, namely $ (0,\dots,0,1)$ with eigenvalue $1$. To diagonalize $A$ thus one needs first to diagonalize $T$. 

The block circulant matrix can be diagonalized by applying a block Fourier transform $F^{\dagger} T F$, where $F$ is a matrix of $n \times n$ blocks of size $n-1 \times n-1$. The block at the $i$-th row and $j$-th column of $F$ is a diagonal matrix with constant elements $\exp( 2 \pi \ii j k /n)/\sqrt{n}$.

After applying this similarity transformation, we will end up with a block diagonal matrix with $n$ blocks, where the $k$-th block is

\begin{equation}
    T_k= \sigma\tau
    \begin{pmatrix}
        3+d_0 & d_1 & d_2 & 0 & 0 & \dots & 0 \\
        d_{-1} & 4+d_0 & d_1 & d_2 & 0 & \dots & 0 \\
        d_{-2} & d_{-1} & 4+d_0 & d_1 & d_2 & \dots & 0 \\
        & \ddots & \ddots & \ddots & \ddots & \ddots & & \\
        0 & \dots & d_{-2} & d_{-1} & 4+d_0 & d_1 & d_2 \\
        0 & \dots & 0 & d_{-2} & d_{-1} & 4+d_0 & d_1 \\
        0 & \dots & 0 & 0 & d_{-2} & d_{-1} & 3+d_0 \\
    \end{pmatrix},
    \label{eq:PBCdiag_block}
    \end{equation}
where $d_{-2} =q^4 \exp(2 \pi \ii k /n)$, $d_{-1} =   2 q^2 (1+\exp(2\pi \ii k /n )) $, $ d_0 = 2\cos(2 \pi k /n)$, $d_1=  2/q (1+\exp(-2 \pi \ii k /n)) $ and $d_2 = 1/q^4 \exp(- 2 \pi \ii k /n)$, $k \in \{1,\dots,n\}$. Following \cite{produkt_matrik} we can rewrite the matrix above as a product of two commuting matrices

\begin{equation}
        T_k = \sigma \tau \tilde{A}_k \tilde{B}_k, \quad \left[\tilde{A}_k,\tilde{B}_k \right] = 0,
   \label{eq:produkt_matrik}
\end{equation}
where 

\begin{align}
    \tilde{A}_k &= \begin{pmatrix}
        a_1 & b_1 & 0 & \dots & 0\\
        c_1 & a_1 & b_1 & \dots & 0\\
        & \ddots & \ddots & \ddots & \\
        0 & \dots & c_1 & a_1 & b_1\\
        0 & \dots & 0 & c_1 & a_1
        \end{pmatrix}, \\
    \tilde{B}_k  &= \begin{pmatrix}
        a_2 & b_2 & 0 & \dots & 0\\
        c_2 & a_2 & b_2 & \dots & 0\\
        & \ddots & \ddots & \ddots & \\
        0 & \dots & c_2 & a_2 & b_2\\
        0 & \dots & 0 & c_2 & a_2
        \end{pmatrix},
    \label{eq:Ts}
\end{align}
where $a_{1} = q^{2} (1+\exp(2 \pi \ii k /n))$, $b_1 = 1$, $c_1 = q^4 \exp(2 \pi \ii k /n)$, $a_2 = q^{-2} (1+\exp(-2 \pi \ii k /n))$, $c_2 = 1$ and $b_2 = q^{-4} \exp(-2 \pi \ii k /n)$. The tridiagonal matrices $\tilde{A}_k$ and $\tilde{B}_k$ can be diagonalized simultaneously. The eigenvalues of $T$ are, after simplifications, 

\begin{equation}
   \lambda_{j,k} = \delta^2 (\cos(\frac{\pi j}{n}) +\cos( \frac{\pi k}{n} ))^2,
    \label{eq:PBC_spekter}
\end{equation}
where $k \in \{1,\dots,n\}$ is the momentum from the Fourier transformation and $j = \{1,\dots,n-1\}$ runs through the components in the $k$-th Fourier mode. The corresponding left and right eigenvectors $(l \cdot n + m)$-th components are

\begin{align}
    \left[r_{j,k}\right]_{l \cdot n + m} &= \frac{\sqrt{2} q^{2 m}}{n} \mathrm{e}^{2 \pi \ii k l/ n} \mathrm{e}^{\pi \ii k m/n} \sin(j m \pi /n), \\
    \left[l_{j,k}\right]_{l \cdot n + m} &= \frac{\sqrt{2} q^{-2 m}}{n} \mathrm{e}^{-2 \pi \ii k l/ n} \mathrm{e}^{-\pi \ii k m/n} \sin(j m \pi /n), 
\end{align}
for $l \in \{1,\dots,n\}$ and $m\in \{1,\dots,n-1\}$.

\subsection{Pseudospectrum}
\label{app:PBC-ps}

\begin{figure}[h]
    \begin{center}
    \includegraphics[width=70mm]{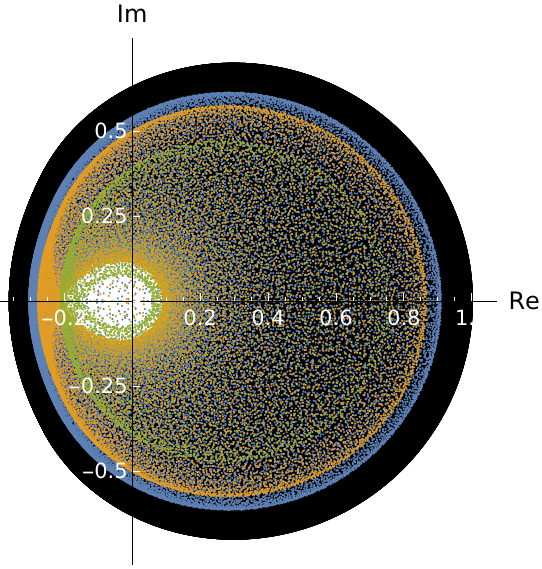}
        \caption{The colored dots represent the $\epsilon$-pseudospectrum for $n=80$ (green), $n=120$ (orange) and $n=180$ (blue) and $\epsilon = 10^{-5}$. As we increase $n$, the $\epsilon$-pseudospectrum starts filling the black region, which is our conjecture for the pseudospectrum of $T$. The black region is obtained as the union of all curves from the conjecture for every possible value of $k/n$.}
    \label{fig:PBC_pse}
    \end{center}
\end{figure}

In this Subsection, we shall compute the pseudospectrum of $T$. In Fig.~\ref{fig:PBC_pse} we plot the $\epsilon$-pseudospectrum for system sizes $80$,$120$ and $180$ and $\epsilon=10^{-5}$. The largest value in the pseudospectrum approaches $1$ as we increase the system size. We conjecture that in the thermodynamic limit the pseudospectrum $T$ is the union over all values $k$ of the product of the pseudospectra of $\tilde{A}_k$ and $\tilde{B}_k$ from Eqs.\ref{eq:Ts}. This coincides with $\left(c_1 \mathrm{e}^{\ii \phi} + a_1 + b_1 \mathrm{e}^{-\ii \phi}\right) \left(c_2 \mathrm{e}^{\ii \phi} + a_2 + b_2 \mathrm{e}^{-\ii \phi}\right) $, for $\phi \in \left[0,2\pi\right]$ and $k/n \in \left[0,1\right]$. The conjectured region is shown in Fig.~\ref{fig:PBC_pse} in black and it seems to match with the plots from Fig.~\ref{fig:PBC_pse} for $n \rightarrow \infty$.

\section{OBC case}
\label{app:OBC}

\subsection{Transfer matrix derivation}
\label{app:OBC-der}

The derivation of the Markov process for OTOC in OBC circuit follows exactly the same rules as the derivation for PBC. In the case of OBC however, we can reduce the dimensionality of our problem if we consider an OTOC between the first qudit and the $j$-th qudit. In this case, the left edge of initial domain will correspond to the left boundary of the system, so the only part of the domain that can move is its right edge. This means that we have to keep track of just $n$ different domain widths.

We begin the derivation of the transfer matrix by encoding all the domains by increasing width in a $n$ dimensional vector $ |v\rangle $. The transfer matrix propagating the domains is 

\begin{equation}
    A = \begin{pmatrix}
            \multicolumn{4}{c}{\multirow{3}{*}{T}} & 0 \\
            \multicolumn{4}{c}{} & \vdots \\
            \multicolumn{4}{c}{} & 0 \\
            0 & \dots & 0 & \sigma & 1
        \end{pmatrix};
    \quad
    T = 
    \begin{pmatrix}
        \delta & \tau & 0 & \dots & 0  \\
        \sigma & \delta & \tau & \dots & 0  \\
        \vdots & \ddots & \ddots & \ddots &  \vdots  \\
        0 & \dots & \sigma & \delta & \tau   \\
        0 & \dots & 0 & \sigma & \delta   \\
   \end{pmatrix},
\label{eq:OBCtransfer_matrix}
\end{equation}
where $\delta = \frac{2 q^2}{(1+q^2)^{2}}$, $\tau=\frac{1}{(1+q^2)^{2}}$ and $\sigma=\frac{q^4}{(1+q^2)^{2}}$. The initial vector must contain the domain on just the first spin, so $|v\rangle = ( \frac{q^4}{q^4-1} ,0,\dots,0)$. The vector used to extract $O(t)$ depends on the qudit $j$ and it is $p_k = 1$ for $k\geq j/2$ and $0$ otherwise.

\subsection{Transfer matrix properties}
\label{app:OBC-properties}

Similarly as for the matrix $A$ for PBC, for OBC we can also compute the spectrum of $A$ by first computing the spectrum of the Toeplitz tridiagonal matrix $T$. Namely, if $\lambda_k$ and $ |v_k\rangle $ are an eigenpair of $T$, then $\lambda_k$ is also an eigenvalue of $A$ whose eigenvector is obtained from $ |v_k\rangle $ by adding one last component $ \frac{ \sigma \left[v_k\right]_{n-1}}{\lambda_k-1} $. There is also an eigenvector of $A$ which is not derived from the eigenvectors of $T$, namely $ (0,\dots,0,1)$ with eigenvalue $1$. The eigenvalues of $T$ are \cite{toeplitz} 

\begin{equation}
    \lambda_k = \delta + 2 \sqrt{\sigma \tau} \cos( \frac{k \pi}{n} ),
    \label{eq:OBC_spek}
\end{equation}
where $k \in \{1, \dots, n-1\}$. The right eigenvectors $ |r_k\rangle $ and left eigenvectors $ \langle l_k|$ are

\begin{align}
    \left[r_k\right]_j &= \left(\frac{\sigma}{\tau}\right)^{j/2} \sin{( \frac{k j \pi}{n})} \\
    \left[l_k\right]_j &= \frac{2}{n} \left(\frac{\tau}{\sigma}\right)^{j/2} \sin{( \frac{k j \pi}{n})}.
\end{align}

\bigskip

The pseudospectrum of $T$ is \cite{trefethen} 

\begin{equation}
    \delta + \mathrm{e}^{\ii \phi} \tau + \mathrm{e}^{- \ii \phi} \sigma, \quad \phi \in \left[0,2\pi\right].
    \label{eq:OBCpseudo}
\end{equation}
Note that the spectrum of $T$ lies on the real values, meanwhile the pseudospectrum is an ellipse in the complex plane with largest value $\delta+\sigma+\tau$, see Fig.~\ref{fig:arbitrary}b.

\subsection{Evaluation of $O(t)$ through spectral decomposition of $T$}
\label{app:OBC-sums}

In this Appendix we will evaluate the expression for OTOC

\begin{equation}
     O(t) = \langle p|T^t|v\rangle,
\label{eq:O_diag}
\end{equation}
$p_k=1$ and $v_k=\delta_{k,1}$. As we shall see, the OTOC behaves as $\sim \lp^t = 1$, but if we subtract the steady state of the whole transfer matrix $A$ the leading term cancel out making $O(t)$ decay as $\lambda_2^t$, so we get $O(t)-O(\infty) \asymp \lambda_2^t$ as expected. Even though $O(t)$ does not exhibit phantom behaviour, the evaluation of $O(t)$ is interesting because the phantom appears in the solution, but it gets cancelled with the steady state that comes from the eigenvalue $\lambda_1=1$. This shows that the physical case of OTOC is special, because the choice of vectors $ \langle p|$ and $ |v\rangle $ makes the phantom cancel exactly with $O(\infty)$. 

We begin by writing Eq.~\ref{eq:O_diag} with the help of the spectral decomposition of $T$

\begin{equation}
    O(t) = \frac{2}{n} \sum_{h=1}^{n-1} \lambda_h^t \sqrt{ \frac{\sigma}{\tau} } \sin{ ( \frac{h \pi}{n} )}
         \sum_{k=1}^{n-1} \left( \frac{\sigma}{\tau} \right)^{k/2} \sin{( \frac{h k \pi}{n} )}
\label{eq:spectral_first}
\end{equation}
where $\lambda_h = \delta+2\sqrt{ \sigma\tau} \cos{(\frac{h \pi}{n})}$. To simplify the expression we shall first evaluate the sum over the index $k$. We get

\begin{equation}
    \sum_{k=1}^{n-1} \left( \frac{\sigma}{\tau} \right)^{k/2} \sin{( \frac{h k \pi}{n} )} = 
    \sqrt{ \frac{\sigma}{\tau} } \frac{\left(1-(-1)^h \left( \frac{\sigma}{\tau} \right)^{ \frac{n}{2}} \right) \sin{\frac{h \pi}{n}}}{1+ \frac{\sigma}{\tau} - 2\sqrt{ \frac{\sigma}{\tau}} \cos{ \frac{h \pi}{n}}}
\label{eq:sum_right}
\end{equation}

Plugging Eq.~\ref{eq:sum_right} into Eq.~\ref{eq:spectral_first} and rearranging some terms we get

\begin{equation}
    O(t) = \frac{2 \tau \left(\frac{\sigma}{\tau}\right)^{\frac{n+1}{2}}}{n}
         \sum_{h=1}^{n-1} \lambda_h^t \sin^2( \frac{h \pi}{n})\frac{\left( \left( \frac{\tau}{\sigma} \right)^{\frac{n}{2}} -(-1)^h\right)}{\lp-\lambda_h},
\label{eq:O_one_sum}
\end{equation}
where we labeled the largest value in the pseudospectrum $\delta+\sigma+\tau $ with $\lp$. To simplify Eq.~\ref{eq:O_one_sum} we will assume $\tau<\sigma$ and $n \gg 1$, so that we can neglect the term $\left( \frac{\tau}{\sigma} \right)^{ \frac{n}{2} }$.

We are interested in the time dependence of $O(t)$, so we can forget about the factors outside the sum, because they do not contribute to the time dependence of $O(t)$. After these simplifications we get

\begin{equation}
    \sum_{h=1}^{n-1} \lambda_h^t (-1)^h \frac{\sin^2\left( \frac{h \pi}{n}\right) }{\lp-\lambda_h}
    \label{eq:sum_1}
\end{equation}

We can replace the fraction $ \frac{1}{\lp-\lambda_h} $ with $ \frac{1}{\lp} \sum_{r=0}^{\infty} \left( \frac{\lambda_h}{\lp} \right)^r$, because $\lp > \lambda_h$.  Omitting constant terms we get

\begin{align}
    &\sum_{r=0}^{\infty} \sum_{h=1}^{n-1} \lambda_h^t (-1)^h \sin^2\left( \frac{h \pi}{n}\right) \left( \frac{\lambda_h}{\lp} \right)^r = \nonumber \\
    &\lp^t \sum_{r=0}^{\infty} \sum_{h=1}^{n-1} (-1)^h \sin^2\left( \frac{h \pi}{n}\right) \left( \frac{\lambda_h}{\lp} \right)^{r+t} = \nonumber \\
    &\lp^t \sum_{k=t}^{\infty} \sum_{h=1}^{n-1} (-1)^h \sin^2\left( \frac{h \pi}{n}\right) \left( \frac{\lambda_h}{\lp} \right)^{k}.
    \label{eq:sum_k}
\end{align}
To prove that $O(t)$ decays as $\lp^t$ we will show that the terms in the sum over the index $k$ are zero for $k \approx n$, meaning that the sum over $k$ runs from $k \approx n$ to $\infty$ and it is independent on $t$. This in turn implies that the only time dependence in the expression Eq.~\ref{eq:sum_k} is $\lp^t$.

We continue by summing over the index $h$

\begin{align}
    \sum_{k=t}^{\infty} &\lp^{-k} \sum_{h=1}^{n-1} \mathrm{e}^{\ii \pi h} (- \frac{1}{4})(\mathrm{e}^{\ii \frac{h \pi}{n}}-\mathrm{e}^{-\ii \frac{h \pi}{n}})^2 (\delta+\sqrt{\sigma \tau} \mathrm{e}^{\ii \frac{h \pi}{n}}+\sqrt{\sigma \tau} \mathrm{e}^{-\ii \frac{h \pi}{n}})^k = \\
    \sum_{k=t}^{\infty} &\lp^{-k} \sum_{r=0}^{k} \binom{k}{r} \delta^{k-r} (\sigma \tau)^{r/2} \sum_{s=0}^{r} \binom{r}{s} \sum_{h=1}^{n} (\mathrm{e}^{\ii \frac{h \pi}{n} (2+r-2s+n)}+\mathrm{e}^{\ii \frac{h \pi}{n} (-2+r-2s+n)} - 2\mathrm{e}^{\ii \frac{h \pi}{n} (r-2s+n)}) = \\
    \sum_{k=t}^{\infty} &\lp^{-k} \sum_{r=0}^{k} \binom{k}{r} \delta^{k-r} (\sigma \tau)^{r/2} \sum_{s=0}^{r} \binom{r}{s} (D_{n-1}(2+r-2s+n)+D_{n-1}(-2+r-2s+n)-2 D_{n-1}(r-2s+n)),
\end{align}
where $D_n(x) = \frac{\sin( (n+1/2) x)}{\sin(x/2)}$ is the Dirichlet kernel \cite{dirichlet}. The Dirichlet kernel is composed of a term $(-1)^{x+1}$, which cancels with the other functions $D_n(x)$ and an infinite series of Kronecker delta functions $\delta_{x/2,\pi p}$, $p \in \mathbb{Z}$. In our case $x= \frac{r-2s+n + 2}{n}$ in the first Dirichlet kernel. The only possible values of $p$ are $0$ and $1$, because for other values there are no $s$ and $r$ to satisfy $x/2=\pi p$. Moreover, for $p=0$ and $p=1$ not all values of $r$ give non-zero contribution. For example $x=r-2s+n+3$ and $p=1$ gives $s=(r-n+1)/2$. The variable $s$ runs from $0$ to $n$, so $(r-n+1)/2 \geq 0$ and $(r-n+1)/2 \leq r$. The former bound gives $r \geq n-1$, the latter $r\geq -n$. Computing these bounds for all three Dirichlet kernels and for both $p=0$ and $p=1$ we see that $r \gtrsim n$ for each term. This in turn implies that the sum over $r$ runs from $\approx n$ to $k$, so $k\gtrsim n$ to have non-zero contributions. For $k<n$ all terms in the sum over $k$ will be zero, so we can substitute $\sum_{k=t}^{\infty}$ with $\sum_{k\approx n}^{\infty}$, which makes the only time dependent part of the leading term in $O(t)$ $\lp^t$.

We saw that $O(t)$ behaves as $\lp^t$, but OTOC decay to $O(\infty)$ as $\lambda_2^t$. This comes from the fact that the leading term in Eq.~\ref{eq:O_one_sum} (the one with $(-1)^h$) exactly sums to $1$ for $t<n$ (not shown)). This means that by subtracting $1$ from $O(t)$ we cancel out the leading term of $O(t)$. The subleading term (the neglected term in Eq.~\ref{eq:O_one_sum}) decays as $\lambda_2^t$, hence OTOC decays as given by the largest eigenvalue of $A$.

\bigskip

The exact evaluation of $ \langle p|T^t|v\rangle $ above can be also used for the case $ p_k = \mu^{-k}$. The choice of $ \langle p|$ reflects in the evaluation of $ \langle p|r_h\rangle $, where $ |r_h\rangle $ is an eigenvector of $T$. We get

\begin{align}
    \langle p|r_k\rangle = \sum_{k=1}^{n-1} \left(\frac{\sigma}{\tau}\right)^{k/2} \mu^{-k} \sin \left( \frac{h k \pi}{n} \right).
\label{eq:Pv_loc}
\end{align}
The Eq.~\ref{eq:Pv_loc} is exactly Eq.~\ref{eq:sum_right} if we substitute $\tau$ with $\tau \mu^2$. We can repeat all calculations from before until we get

\begin{equation}
    O(t) \propto \sum_{h=1}^{n-1} \lambda_h^t (-1)^h \frac{\sin^2 ( \frac{h \pi}{n+1}) }{\lambda(\mu)-\lambda_h},
\label{eq:lambda_arb}
\end{equation}
with $\lambda(\mu) = \delta + \sigma/\mu + \tau \mu$. Eq.~\ref{eq:lambda_arb} is analogous to Eq.~\ref{eq:sum_1} if we replace $\lambda(\mu)$ with $\lp$. We can repeat all calculations below Eq.~\ref{eq:sum_1} to show that $\lambda(\mu)$ is indeed the true decay of $O(t)$. The value of $\lambda(\mu)$ is greater than $\lambda_2$, because otherwise we cannot repeat the steps in Eq.~\ref{eq:sum_k}. Moreover, $\lambda(\mu) \leq \lp=1$ otherwise $ \langle p|$ is not normalizable. We conclude that $\lambda_2 \leq \lambda(\mu) \leq \lp$.

\subsection{Biased random walk}
\label{app:OBC-randomwalk}

In this Section we prove that OTOC dynamics in a OBC BW circuit is equivalent to a biased 1D random walk coupled to reservoirs at the edges. Using this equivalence, we will be able to compute the decay of OTOC to $O(\infty)$, namely $O(t)-O(\infty) \asymp \lambda_2^t$, where $\lambda_2$ is the second largest eigenvalue of the transfer matrix that propagates OTOC (see App.~\ref{app:OBC-properties}). 

\begin{figure}[h]
    \begin{center}
    \includegraphics[width=100mm]{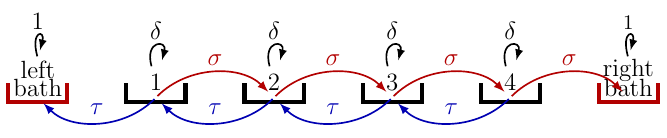}
        \caption{Cartoon picture of a biased random walk on $4$ sites with dissipation on the boundary. Above the sites (boxes) are shown possible moves (arrows) with the corresponding probability. One the random walk enters the reservoirs it cannot return back to the bulk of the chain.}
    \label{fig:random_walk}
    \end{center}
\end{figure}

The transfer matrix from Eq.~\ref{eq:transfer_matrix}, used to propagate OTOC, can be also used to describe a 1D biased random walk. To get a proper Markov chain transfer matrix, the elements in each column should sum to $1$ to conserve probabilities. To achieve this and keep the tridiagonal transfer matrix to propagate probabilities one can make the random walk dissipate on the left and right boundary, as shown in Fig.~\ref{fig:random_walk}. Doing so, the Markov chain transfer matrix is 

\begin{equation}
    A = 
    \begin{pmatrix}
        1 & \tau & 0 & 0 & 0 & \dots & 0 \\
        0 & \delta & \tau & 0 & 0 & \dots & 0  \\
        0 & \sigma & \delta & \tau & 0 & \dots & 0 \\
        \vdots & \vdots & \ddots & \ddots & \ddots &  & \vdots  \\
        0 & 0 & \dots & \sigma & \delta & \tau & 0  \\
        0 & 0 & \dots & 0 & \sigma & \delta & 0  \\
        0 & 0 & \dots & 0 & 0 & \sigma & 1 
   \end{pmatrix},
\label{eq:Markov_transfer_matrix}
\end{equation}
where $\tau + \sigma + \delta = 1$, and where the first and last site correspond to the left and right bath, respectively. Note that we could propagate OTOC using this transfer matrix, with $p_k = 1-\delta_{k,1}$ and $v_k=\delta_{2,k}$. Because $p_1=0$, the OTOC can be equivalently propagated by forgetting about the first column and row in Eq.~\ref{eq:Markov_transfer_matrix}. Using the random walk interpretation, the OTOC $O(t) = \langle p|A^t|v\rangle $ is equivalent to the probability of not being in the left bath by beginning at the second position (left edge of the bulk).

Let us label the probability of being at site $j$ in the bulk with $r_{j+1}$ (so the leftmost site has probability $r_2$), and the probability of being in the left bath with $r_1$, then OTOC is $O(t) = 1-r_1$. To prove that OTOC decays to its asymptotic value $O(\infty)=1$ as $\lambda_2^t$, we will compute the probability $r_1$. Because we can enter the left bath just once, the probability $r_1$ of being in the left bath at time $t$ can be found by summing the probabilities of entering the left bath for all times $T+1 \leq t$. The probability of entering the left bath at time $T+1$ is obtained as $\tau$ times the probability of being at the leftmost site in the bulk $1$ at time $T$ without ever being in the left reservoir before. The number of paths of length $T$ beginning at $1$ and ending at $1$ without going to the left reservoir corresponds to the number of Dyck words \cite{dyck} with $\sigma$ and $\tau$. The number of these Dyck words can be expressed with the Catalan number $C_{T/2}$ \cite{catalan}, which is $\binom{T}{T/2}/(T/2+1)$ if $T$ is even and $0$ otherwise. The corresponding probability for such a Dyck word is $(\tau \sigma)^{T/2}$. To get the probability of being at $1$ at time $T$ without ever being in the left bath, we should also include the moves where we stay at the same place; this moves have probability $\delta$ and can be places anywhere in the Dyck word. To sum up, the probability of being at site $1$ at time $T$ is obtained as the sum over all possible combinations of moves to left/right and staying at the same position. At the end we get

\begin{equation}
    r_1(t) = \tau \sum_{T=0}^{t-1} \sum_{k=0}^{T/2} \frac{\binom{2k}{k}}{k+1} (\tau \sigma)^{k} \delta^{T-2k} \binom{T}{2k}.
\label{eq:OTOC_a}
\end{equation}
OTOC is obtained as $1-r_1$. For simplicity, we will now compute $O(t)$ for qubits, $q=2$. We get

\begin{align}
    O(t) &= 1-r_1(t) \nonumber \\
         &= 1+ \frac{64}{375 \sqrt{\pi}} ( \frac{16}{25} )^t
         \frac{\Gamma(3/2+t)}{\Gamma(3+t)} {}_2 F_1(1,3/2+t,3+t,16/25),
\label{eq:OTOC_closed}
\end{align}
where ${}_2 F_1(a,b,c;z)$ is the hypergeometric function. With the help of the biased random walk picture, in Eq.~\ref{eq:OTOC_closed} we obtained a closed form solution of OTOC in OBC BW circuits with Haar random 2-site gates. A closed form solution was obtained also in \cite{PRR}, but it is much more complex, because it is expressed with recursion. In \cite{adam18} there is a simple result for OTOC, but it is for infinite systems, whereas our solution holds for any system size $n$.

\begin{figure}[H]
    \begin{center}
    \includegraphics[width=90mm]{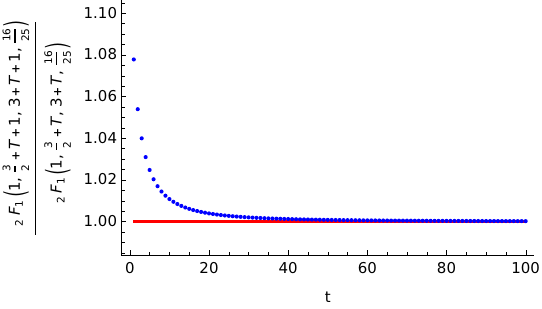}
        \caption{Ratio of the two hypergeometric functions from Eq.~\ref{eq:rate}. The ratio decays fast to $1$.}
    \label{fig:ratio}
    \end{center}
\end{figure}

Because of the simple form of Eq.~\ref{eq:OTOC_closed}, we can compute the rate at which OTOC decays to $O(\infty)$. Assuming OTOC decays to the asymptotic value exponentially, we can get the exponent as $\leff(t) = \frac{O(t+1)-1}{O(t)-1} $. We get

\begin{equation}
    \leff(t) = \frac{16}{25} \frac{3/2 +t}{3+t} \frac{{}_2 F_1(1,5/2+t,4+t,16/25)}{{}_2 F_1(1,3/2+t,3+t,16/25)}.
\label{eq:rate}
\end{equation}
The term $\frac{3/2 +t}{3+t}$ behaves as $1- \frac{3}{2t} + O(1/t^2)$, and the ratio of the hypergeometric functions decays to $1$ faster than exponentially (Fig.~\ref{fig:ratio}), so we conclude that $O(t)$ decays to $1$ as $\leff = 16/25 = \lambda_2$.

We computed the time evolution of a random walk, where we start at the leftmost position in the bulk and we are interested in the probability of staying in the bulk. This random walk coincide with the OTOC evolution in OBC random quantum circuits. When computing the relaxation of OTOC to their asymptotic value $O(\infty)$, we subtract the leading term from $O(t)$, which results in the relaxation given by $\lambda_2$. For different initial conditions subtracting $O(\infty)$ from $O(t)$ does not cancel exactly the leading term in $O(t)$, meaning that $O(t)$ initially does not relax, but stays constant, as we expect by looking at the largest value in the pseudospectrum of the transfer matrix, $\lp=1$. OTOC in this sense can be considered a special case of initial conditions.

\section{Hermitian transfer matrix}
\label{app:herm}

\begin{figure}[h]
    \begin{center}
    \includegraphics[width=70mm]{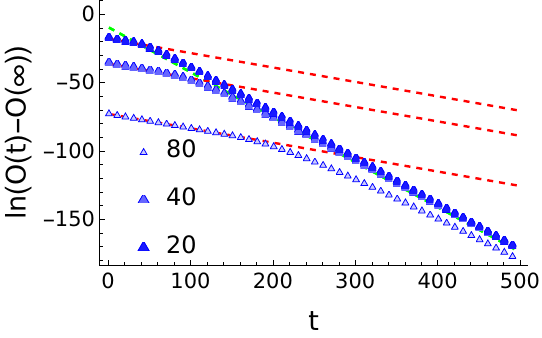}
        \caption{Iteration of the tridiagonal matrix $T$ with $\delta=8/25$, $\tau=\sigma=5/25$ for $p_k\propto \mu^{-k}$ and $v_k=\delta_{k,1}$, $\mu=0.4$. The vector $ \langle p|$ is normalized such that $\sum_k p_k=1$. The quantity $ \langle p|T^t|v\rangle$ decays towards its asymptotic values as $\lph^t \approx 0.85^t$, and $\lambda_2 \approx 0.72$. The red line denotes $\lph \approx 0.85$ and the green line the asymptotic decay of finite system sizes $\lambda_2$. Even though the initial decay $\lph$ is larger than $\lambda_2$, the plot for different system sizes clearly shows that $O(t=0)$ approaches $0$ by increasing the system size $n$.}
    \label{fig:hermitian}
    \end{center}
\end{figure}

In this Appendix, we shall prove that the phantom decay $\lph > \lambda_2$ of quantities $O(t) = \langle p|T^t|v\rangle $ to their asymptotic value $O(\infty)$ is a finite size effect if $T$ is a symmetric tridiagonal Toeplitz matrix with upper and lower diagonal elements $\beta$. Increasing the system size, $O(t)$ approaches $0$ for every non-zero time as seen if Fig.~\ref{fig:hermitian}. In the TDL, $O(t)$ will be exactly equal to zero for every non-zero time.

Let us begin with the initial vectors $ p_k \propto \mu^{-k}$ and $v_k = \delta_{k,1}$. The expression $O(t)$ can be computed using $p_k = 1$ if we appropriately transform $T$ with a similarity transformation, similarly as we did in the main text in Eq.~\ref{eq:tridiagonal_rescaled}. In this case we get an effective iteration with a non-Hermitian matrix, where a decay $\lph > \lambda_2$ should not be surprising. We have

\begin{equation}
    D^{-1} T D = \begin{pmatrix}

        \delta & \beta \mu & 0 & \dots & 0 \\
        \frac{\beta}{\mu}& \delta & \beta \mu & \dots & 0 \\
        \vdots & \ddots & \ddots & \ddots & \vdots \\
        0 & \dots & \frac{\beta}{\mu} & \delta & \beta \mu \\
        0 & \dots & 0 & \frac{\beta}{\mu} & \delta
        \end{pmatrix}
\label{eq:app-tridiagonal_rescaled}
\end{equation}
where $D$ is a diagonal matrix with diagonal elements equal to $D_{k,k} = \mu^k$. The choice $\mu > 1$ is the only possible choice if we wish to normalize $ \langle p|$. In this case the upper diagonal of $D^{-1} T D$ is larger than its lower diagonal, resulting in a transfer matrix of the same form as the one used to propagate OTOC. Even the initial vectors $p_k = 1$ and $v_k = \delta_{k,1}$ are the same as those used for OTOC, so we know that $O(t)$ will decay to its asymptotic value $O(\infty) = 1 $ as $\lambda_2$.

Another possible choice for $ \langle p|$ is when $\mu < 1$. In this case $\lph > \lambda_2$, however if we wish to normalize $ \langle p| $, then $O(t=1)$ will scale as $ \langle p|v\rangle = \frac{\mu^{-1}}{\sqrt{\sum_k \mu^{-2k}}} \approx \mu^{-n}$. The normalization and right-localization of $ \langle p|$ implies that $O(t)$ is exactly zero for every non-zero time in the TDL.

To conclude, we explored the case of Hermitian $T$ and saw that the rate at which $O(t)$ decays to $O(\infty)$ is $\lph>\lambda_2$ only $ \langle p|$ localized at the right edge of the system, i.e. $\mu<1$. In this case, however, $O(t)$ decays to $0$ for every non-zero time as we increase the system size $n$, which makes the phantom eigenvalue a finite size effect in Hermitian systems. In contrast, when $T$ is non-Hermitian, $\lph > \lambda_2$ until extensive times also when $O(t)$ is not zero in the TDL, which makes $\lph$ the only true decay in the TDL.

\end{document}